\documentstyle[editedbook,epsfig,psfig,epsf]{mq}
\def\etal{{\em et al.} }

\def\cm2{cm$^2$ }
\def\se1{s$^{-1}$ }

\newcommand{\grs}{GRS 1915+105 }

\begin{opening}
\title{GMRT radio observations of microquasar GRS 1915+105}
\author{J. S. Yadav$^1$, C. H. Ishwara-Chandra$^2$, A. Pramesh
Rao$^2$ and G. G. Pooley$^3$}
\institute{$^1$ TIFR, Homi Bhabha Road, Mumbai - 400 005 INDIA \\
$^2$ NCRA, Post Bag No. 3, Ganeshkhind, Pune - 411 007 INDIA  \\
$^3$ Mullard Radio Astronomy Observatory, Cambridge, UK\\}
\end{opening}

\runningtitle{Workshop Proceedings}
\runningauthor{Yadav et al.}

\begin{document}
\vspace{-0.5cm}
\begin{abstract}
{\small We have observed  microquasar \grs at 1.28
GHz for 8 days from June 18 to July 1, 2001 using Giant Metrewave  Radio
Telescope (GMRT). We have seen several isolated radio baby flares of 
varying intensity and  duration.  We have also observed  broad composite flares
with  rise and decay times of few hours on June 28-29, 2001 few days
prior to when source went to the ``plateau radio state'' on 3rd July, 2001.
These broad radio flares consist of several overlapping baby radio flares.
The source was in the low-hard X-ray state during this period. 
We compare these results with 15 GHz radio data from the Ryle telescope.}
\end{abstract}

\section{Introduction}
The long term  monitoring of microquasar  \grs has  shown        
broad  correlation between the   X-ray and non-thermal radio emission  
\cite{herm97}. The radio emission in \grs can be  classified into
three classes; (i) the relativistic superluminal radio jets of flux
density $\sim$ 1 Jy with decay time-scales of several days (\cite{mira94},
\cite{fend99}), (ii) the baby jets of 20 $-$ 40 min durations
with flux density of 20 $-$ 200 mJy both in infrared (IR) and radio
(\cite{pool97}, \cite{eike98}), and (iii) the plateau
 state with persistent radio emission of 20 $-$ 100 mJy for extended
durations \cite{muno01}. In the case of superluminal jets, the radio
emission has steep spectra and are observed at large distances (400 $-$
5000 AU) from the accretion disk \cite{fend99}, \cite{dhaw00}.
The radio emission during other two classes has flat spectra and they occur
close to the accretion disk (within a few tens of AU)\cite{dhaw00}. 
The multi-wavelength
studies have indicated strong disk-jet connection for class 2(\cite{mira98},  
\cite{eike98}, \cite{yada01}).   In this paper, we present our radio
observations at 1.28 and 15 GHz during June 18 to July 3, 2001 (just prior  
to when  source went  to the plateau state on July 3). These observations
include faint radio emission, isolated baby radio flares riding above
smooth as well as slowly rising/decaying radio emission (broad flares), 
and large radio
flare ($\sim$ 200 mJy on June 20).
\begin{figure}[h]
\centering
\psfig{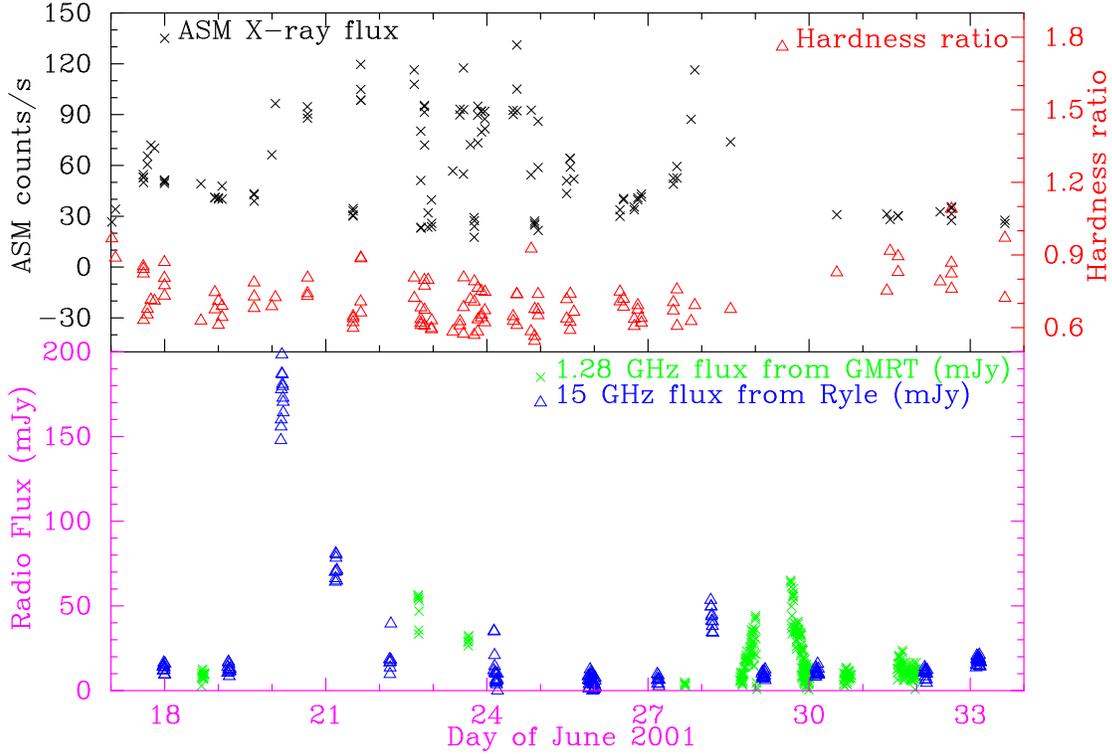}
\caption{The RXTE/ASM flux (left axis) and hardness ratio (right axis) during
June 17 to July 3, 2001(top panel). The observed 1.28 GHz flux from 
GMRT and 15 GHz flux from Ryle telescope are shown in bottom panel 
(5min bin) for the same duration.}
\label{fig:ex}
\end{figure}

\section{Observations and discussion}
The 1.28 GHz radio observations were carried out  with a
 bandwidth of 16 MHz using the GMRT at Pune, India \cite{swar91} 
 on 2001 June
18, 22, 23, 27, 28, 29, 30 and July 1.  The flux density scale is
set by observing primary calibrator 3C286 or 3C48.   
The integration time of 32 s was chosen. The data recorded from
GMRT has been converted to FITS and was analysed using Astronomical
Image Processing System ({\tt AIPS}).  The details 
of observations/analysis is  given elsewhere \cite{ishw02}. 
The radio observations
at 15 GHz are from the Ryle telescope at Cambridge. Details of  
the instrument and analysis  can be found  elsewhere \cite{pool97}.

The RXTE/ASM flux and X-ray hardness ratio (5-12
keV/1.5-5 keV) are  shown in the top panel of Figure 1.
The radio lightcurves at 1.28 and 15 GHz produced at 5 min interval are 
shown in the bottom panel of
Figure 1.  \grs exhibited significant radio emission on all days,
except on June 27,  when the source showed much weaker radio emission 
($\sim$ 5 mJy). Figure 1 brings out following important points:
\begin{enumerate}
\item The radio fluxes at 1.28 and 15 GHz are consistent and suggest a 
flat radio spectrum during these observations except during the large radio
flare on June 20-24. The flux and decay time of this flare suggest that it
belongs to the class of relativistic jets. The higher flux at 1.28 GHz is
consistent with steep spectrum during the decay of this flare.
\item The X-ray hardness ratio changed from $\sim$ 0.7 to $\sim$ 0.85 
around June 30 prior to when source went  to the plateau state on July 3.
The RXTE/ASM flux  becomes stable at higher hardness ratio. The X-ray
hardness ratio suggests that the source remains in the low-hard state
during this period.
\end{enumerate}

\begin{figure}[t]
\centering
\psfig{file=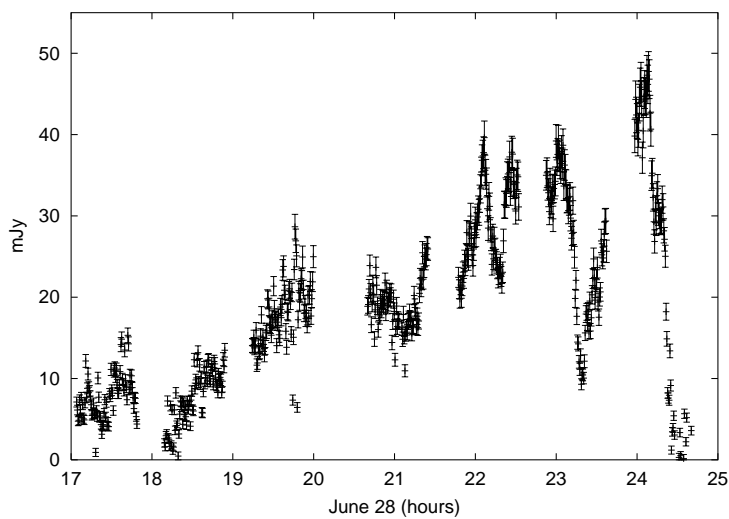,width=7cm,angle=270}
\caption{The radio light curve from GMRT at 1.28 GHz (mJy) observed on June 28, 2001 (5 min. bin).}
\label{fig:ex}
\end{figure}
The radio light curve observed on June 28 is shown in Figure 2. 
The flux density was $\sim$ 5 mJy at about
UTC 17 hour and reached gradually to 50 mJy at UTC 24 hour. When the
observations resumed on 29 (not shown here), 
the source was "caught" at 70 mJy at UTC
16 hour, and the flux started decaying slowly to the value of 10 mJy at
UTC 24 hour. This is probably the first detail observation of the
precursor flares to the plateau state with rise and decay times of 
over  six hours. It is also important to note that these broad flares 
and the change in the hardness ratio around June 30 set the stage for
the plateau state.

These broad flares   consist of mini (baby) radio flares 
riding above rising/decaying radio
emission. In contrast, the radio lightcurve observed on June 30 shows
mini (baby) flares riding above almost smooth radio emission (Figure 3).

\begin{figure}[h]
\centering
\psfig{file=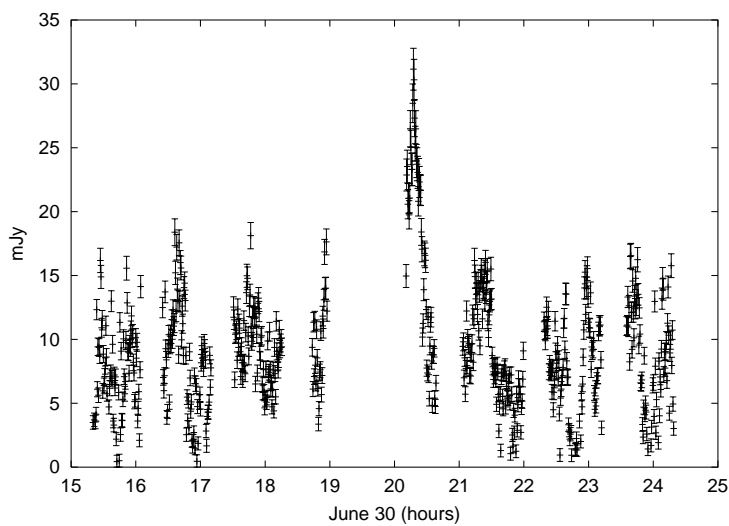,width=7cm,angle=270}
\caption{The radio light curve from GMRT at 1.28 GHz (mJy)observed on June 30, 2001 (
5 min. bin).}
\label{fig:ex}
\end{figure}

These isolated baby flares are modeled as adiabatically expanding 
synchrotron clouds ejected from the accretion disk \cite{ishw02}. One
such flare along with model fit data is shown in Figure 4. This model 
provides estimate of the spectrum index from single frequency observations
and successfully explains the observed delay times between different 
frequencies.
The radio emission during these observations is consistent with flat radio
spectrum which is in agreement with observed flux at 15 GHz.
\begin{figure}[h]
\centering
\psfig{file=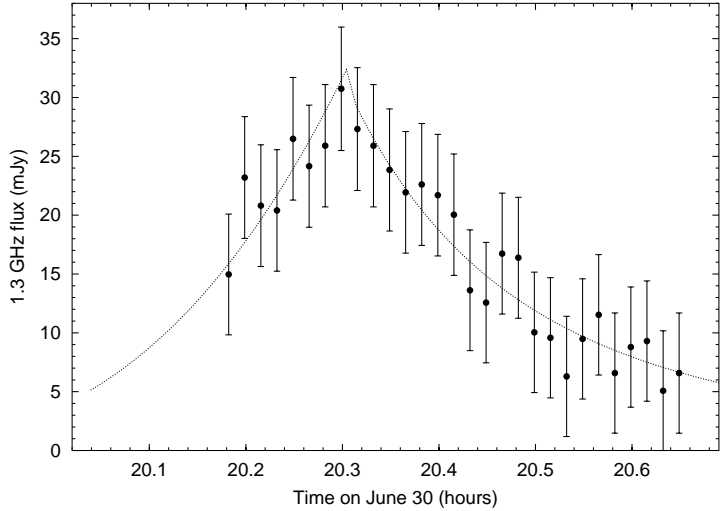,width=7cm,angle=270}
\caption{Isolated radio flare observed with GMRT at 1.28 GHz on 30 June, 2001 with model fit (for details see text).}
\label{fig:ex}
\end{figure}

\section*{Acknowledgments}
We thank the staff of the GMRT that made these observations possible. GMRT
is run by the National Center of Astrophysics of the Tata Institute
of Fundamental Research. We also thank RXTE/ASM team for making 
their data public.

\end{document}